\documentclass[aps,prb,twocolumn,groupedaddress,showpacs]{revtex4} 

\usepackage{graphicx}
\usepackage{dcolumn}
\usepackage{bm}
\usepackage{amssymb}

\begin{document}

\title{Strong charge fluctuations manifested in the high-temperature 
Hall coefficient of high-$T_c$ cuprates}

\author{S. Ono} 
\author{Seiki Komiya} 
\author{Yoichi Ando} 
\altaffiliation{Corresponding author. e-mail: ando@criepi.denken.or.jp}
\affiliation{Central Research Institute of Electric Power 
Industry, Komae, Tokyo 201-8511, Japan} 

\date{\today}

\begin{abstract}

By measuring the Hall coefficient $R_H$ up to 1000 K in La$_2$CuO$_4$,
Pr$_{1.3}$La$_{0.7}$CuO$_4$, and La$_{2-x}$Sr$_x$CuO$_4$ (LSCO), we
found that the temperature ($T$) dependence of $R_H$ in LSCO for $x$ = 0
-- 0.05 at high temperature undoubtedly signifies a gap over which the
charge carriers are thermally activated, which in turn indicates that in
lightly-doped cuprates strong charge fluctuations are present at high
temperature and the carrier number is not a constant. At higher doping
($x$ = 0.08 -- 0.21), the high-temperature $R_H(T)$ behavior is found to
be qualitatively the same, albeit with a weakened temperature
dependence, and we attempt to understand its behavior in terms of a
phenomenological two-carrier model where the thermal activation is
considered for one of the two species. Despite the crude nature of the
model, our analysis gives a reasonable account of $R_H$ both at high
temperature and at 0 K for a wide range of doping, suggesting that
charge fluctuations over a gap remain important at high temperature in
LSCO deep into the superconducting doping regime. Moreover, our model
gives a perspective to understand the seemingly contradicting
high-temperature behavior of $R_H$ and the in-plane resistivity near
optimum doping in a consistent manner. Finally, we discuss possible
implications of our results on such issues as the scattering-time
separation and the large pseudogap.

\end{abstract}

\pacs{PACS numbers: 74.25.Fy, 74.72.-h, 74.72.Dn}

\maketitle 

\section{INTRODUCTION}

The Hall coefficient $R_H$ is generally a useful tool to characterize
metals and semiconductors, for it reflects the band structure and the
sign of charge carriers. In strongly-correlated electron materials,
$R_H$ often shows complex dependences on temperature and other control
parameters, which are expected to give a clue to understanding the
underlying unconventional electronic states; in this context, $R_H$ has
been used for the studies of such novel phenomena as quantum phase
transition \cite{Balakirev,Paschen,Dagan,Yeh} or charge-stripe
formation. \cite{Noda} Nevertheless, quantitative understandings of
$R_H$ in strongly-correlated electron materials have been difficult to
achieve, and high-$T_c$ cuprate superconductors are no exception.
\cite{Orenstein} In fact, since the optimally-doped cuprates are just in
the middle of the strong- and weak-coupling limits,
\cite{Orenstein,Kivelson1} even the basic framework for the description
of its electronic state is still controversial; as a result, most of the
normal-state transport properties, including the Hall coefficient, are
generally believed to be too complicated to be quantitatively
understood, though there have been attempts to explain their fundamental
behavior in frameworks of specific theories.
\cite{Anderson,Stojkovic,Kontani,Veberic}

In this regard, however, the $R_H$ of high-$T_c$ cuprates appears to be
rather peculiar, because in the lightly hole-doped regime where the
low-energy physics is apparently governed by the ``Fermi arc" (a small
portion of the large Fermi surface expected in the absence of electron
correlations),\cite{Orenstein,ARPES} we have recently found \cite{Ando}
that $R_H$ simply behaves like the ``Hall constant" of a conventional
metal and the electron correlation effects seem to be implemented only
through modifications of the Fermi surface in determining $R_H$. This
finding in cuprates suggests that $R_H$ might give us a unique
opportunity to find a clue to building a proper framework for the
description of the electronic state of cuprates despite the strong
electron correlations. It should be emphasized that there is still no
proper theoretical understanding for {\it how} the correlation gap is
suppressed (and eventually closed) when carriers are doped to a Mott
insulator, so any new information regarding the band structure of doped
Mott insulators (to which cuprates belong) would be of fundamental
importance for addressing the strong electron correlation problem.

The La$_{2-x}$Sr$_x$CuO$_4$ system \cite{Kastner} we study here is a
prototypical cuprate, where $x$ = 0 is a parent charge-transfer (CT)
insulator \cite{ARPES,Dagotto,Geballe} with the band gap $\Delta_{CT}$
called the CT gap. We also study another parent CT insulator
Pr$_{1.3}$La$_{0.7}$CuO$_4$ (PLCO) for comparison. In our previous work,
\cite{Ando} besides showing that $R_H$ of LSCO at low doping can be
interpreted as a ``Hall constant" below 300 K, we reported that a marked
decrease sets in at higher temperature, which was proposed to be
possibly due to the contribution of thermally-created holes. In the
present work, in an effort to quantitatively understand the temperature
dependence and the doping evolution of $R_H$ in LSCO and to elucidate
the electronic structure of this system, we have extended the
high-temperature measurements of $R_H(T)$ to a wide doping range,
starting from $x$ = 0. As a result, we find that the behavior of
$R_H(T)$ at $x$ = 0 is surprisingly simple to understand, despite the
strong electron correlations that open the Mott gap in the first place;
furthermore, our data in the lightly-doped region are found to give a
solid basis for understanding the high-temperature behavior in terms of
the thermal activation of charge carriers over a well-defined gap. At
higher doping, the behavior of $R_H(T)$ becomes less unambiguous to
understand, and in this paper we present our attempt to describe its
behavior in terms of a phenomenological two-carrier model.

\section{EXPERIMENTAL}

The La$_{2-x}$Sr$_x$CuO$_4$ and Pr$_{1.3}$La$_{0.7}$CuO$_4$ single
crystals are grown by the traveling-solvent floating-zone method. The
LSCO crystals are annealed to tune the oxygen content to 4.000$\pm$0.001
as described in Refs. \onlinecite{Komiya1} and \onlinecite{Komiya2}. In
particular, the oxygen stoichiometry of the $x$ = 0 samples (which are
annealed in flowing Ar at 800$^\circ {\rm C}$) is confirmed by both the
thermo-gravimetry analysis and the magnetic susceptibility, which shows
a very sharp N\'{e}el transition at 321 K. The $x$ values shown here are
determined by the inductively-coupled-plasma
atomic-emission-spectroscopy analysis. The PLCO crystals are annealed in
flowing O$_2$ at 700$^\circ {\rm C}$ to minimize oxygen deficiency. To
measure $R_H$ at temperatures up to 1000 K, we use a 6-T superconducting
magnet system with a room-temperature bore, in which a water-cooled
furnace is fit. Samples are kept under suitable atmospheres during the
experiments to ensure that the oxygen content does not change. For the
measurements of $R_H$ using a standard six-probe technique, we record
the full magnetic-field dependence of the Hall voltage at a fixed
temperature by sweeping the magnetic field to both plus and minus
polarities, subtract the asymmetric part coming from contact
misalignment, and fit the symmetrical part with a straight line to
determine $R_H$, which is essential for the high accuracy achieved in
the present work. 

Because the present work involves transport measurements in temperatures
up to 1000 K which is much higher than usual, it would be prudent to
discuss the possibility that the high-temperature Hall data might be
adversely affected by the oxygen mobility within a sample or by some
change in the oxygen content: As for the oxygen mobility at high
temperature, this can in principle provide a parallel channel for charge
transport; such an effect is actually important in a mixed ionic and
electronic conductors such as GdBaMn$_2$O$_{5+x}$ which was recently
found to have a very high oxygen mobility,\cite{Taskin_APL} but in LSCO
the oxygen-ion conductivity at 1000 K is negligible compared to the
electron conductivity. Of more concern might be that the moving oxygen
could affect the electron transport in an unexpected way; however, this
is very unlikely to be the case in LSCO, because the oxygen diffusion
time at 1000 K is of the order of 10$^{-8}$ s, while the electron
scattering time is of the order of 10$^{-13}$ s at best, so the
electrons only see static oxygen environment during the transport
events. Regarding the oxygen content issue, it should be emphasized that
in the present experiment the atmosphere in the furnace is optimized for
different doping ranges to minimize the change in the oxygen content in
the samples during the measurements. For example, in the lightly-doped
region of LSCO one should only worry about excess oxygen, so the
measurements are done in pure argon; on the other hand, in the
optimally- to overdoped region one should worry about oxygen vacancies
rather than excess oxygen, so the measurements are done in 1 atm of
oxygen. (Since we cannot avoid noticeable oxygen loss at high
temperature for $x >$ 0.20, we show data only up to 700 K for $x$ = 0.21
in this work.) Since the oxygen phase diagram is known for the whole
range of Sr doping for LSCO,\cite{Kanai} we can be sure that the
possible change in the effective hole doping is less than 1\% during the
measurements up to 1000 K under our measurement conditions. Indeed, the
resistivity data monitored upon thermal cycling (warming to 1000 K and
cooling to room temperature) are always confirmed to be essentially
identical.

\section{RESULTS}

\begin{figure}
\includegraphics[clip,width=8.5cm]{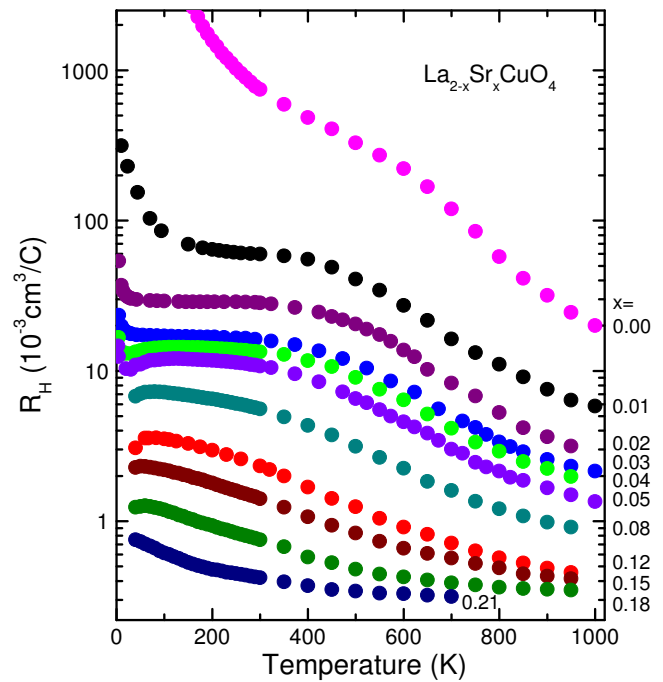} 
\caption{(color online) 
Temperature dependences of $R_H$ up to 1000 K measured on a series of 
high-quality LSCO single crystals.} 
\end{figure} 

Figure 1 shows the $R_H(T)$ data of LSCO single crystals up to 1000 K
for the whole doping range studied. Those data are essentially
consistent with the previously-reported high-temperature data measured
on polycrystalline samples,\cite{Nishikawa,Hwang} but bear much higher
accuracies. Note that there has been no high-temperature data for $x$ =
0 -- 0.03 even for polycrystalline samples, and it turns out that the
data from this insulating regime provide important insights for
analyzing the data for higher dopings. In the following, we describe our
analyses of the data, starting from the parent insulator, $x$ = 0.

\subsection{Parent Insulators}

Interestingly, the temperature dependence of $R_H$ is found to be most
unambiguously understood in the parent CT insulator La$_2$CuO$_4$, where
the electron correlation effects are the strongest; Fig. 2(a) shows the
$R_H(T)$ data up to 1000 K together with the fitting to the formula,
\begin{equation} 
R_H(T) = \frac{V_{Cu}}{e} \left( n_0 e^{-\Delta_{imp}/2k_{B}T} + 
n_1 e^{-\Delta_{CT}/2k_{B}T} \right)^{-1}.
\end{equation} 
It is clear in Fig. 2(a) that this fitting is essentially
perfect, which means that $R_H(T)$ at $x$ = 0 is simply governed by two
gaps ($e$ is the electron charge and $V_{Cu}$ is the unit volume per
Cu). The lower-temperature part of the data tells us that there is a
small concentration ($n_0$ = 0.57\%) of impurity states located at
$\Delta_{imp}$ = 0.087 eV above the top of the valence band; such a
situation has been already known \cite{Kastner} for La$_2$CuO$_4$, where
$\Delta_{imp}$ determines the impurity-ionization energy.\cite{note_imp}
What is new here is that at high temperature above $\sim$500 K another
activation process with $\Delta_{CT}$ = 0.89 eV involving a large
density of states becomes active, which obviously corresponds to
the activation across the CT gap. A schematic energy diagram is shown in
the inset of Fig. 2(a). The fact that $R_H$ is positive at high
temperature means that the holes in the valence band are more mobile
than the electrons in the conduction band; indeed, if one remembers
\cite{Geballe} that the valence band in cuprates is primarily of O 2p
character (implying a wide band), while the conduction band is
essentially a Cu 3d band (which is always narrow \cite{Adler}), it is
reasonable for the holes to have higher mobility and become dominant.
Our fitting yields the prefactor of the high-temperature excitation
term, $n_1$, of as large as 4.3 hole/Cu; although this exceeds the
logical limit of 1 hole/Cu, this is actually reasonable because the
contribution of the electrons (that are created with holes) tends to
reduce the $R_H$ value and lead to an increase \cite{Adler} in the
effective $n_1$.\cite{note_intrinsic} In any case, $n_1$ being larger
than 1 hole/Cu strongly suggests that we are observing excitations
across major bands.

\begin{figure}
\includegraphics[clip,width=8.5cm]{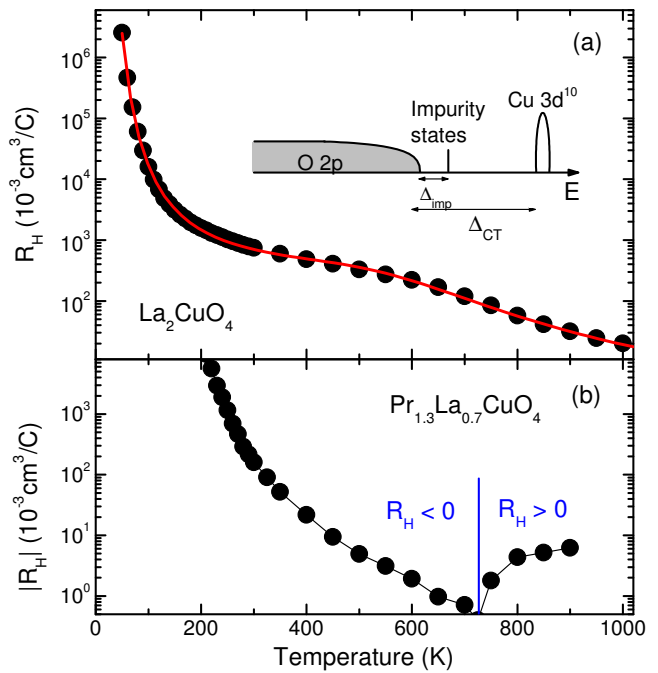} 
\caption{(color online) 
(a) $R_H$ of La$_2$CuO$_4$ measured up to 1000 K. The
solid line is the fit to the data with Eq. (1) with the parameters 
$n_0$ = 0.0057, $\Delta_{imp}$ = 0.087 eV, $n_1$ = 4.3, and 
$\Delta_{CT}$ = 0.89 eV. 
The inset shows a schematic energy diagram for the relevant
bands and states. (b) Absolute value of $R_H$ of
Pr$_{1.3}$La$_{0.7}$CuO$_4$, which shows a sign change
near 700 K.} 
\end{figure} 

Figure 2(b) shows similar data for PLCO, which has the so-called $T'$
structure \cite{Dagotto} and is the parent insulator for an
electron-doped material. The sign of $R_H$ is negative below 700 K,
which indicates that the impurity states (that are presumably due to
some oxygen nonstoichiometry) lie close to the conduction band and
provides electron carriers at moderate temperature. However, $R_H$ of
PLCO shows a sign change above 700 K, corroborating the conclusion
obtained for La$_2$CuO$_4$ that at high temperature the activation
across the CT gap becomes relevant and the hole carriers govern $R_H$ 
due to their higher mobility. 

We note that our $\Delta_{CT}$ obtained for La$_2$CuO$_4$ gives the
thermodynamically measured CT gap, and this is to be compared with the
optically measured CT gap, \cite{ARPES,Kastner} which is usually
believed to be about 2 eV. Here, a little caution is needed, because it
is customary to take the {\it peak energy} of the optical absorption as
a measure of the CT gap, but this energy does {\it not} correspond to
the true gap that is best measured as the minimum excitation energy
between the two bands and is equal to the distance between the two band
edges. When we take the {\it onset} of the optical absorption as a
measure of the CT gap, the existing optical data \cite{Kastner} give
values of 0.9 -- 1.3 eV, which is still a bit larger than our
thermodynamic value of 0.89 eV. An obvious reason for this difference is
that the CT gap is an indirect one, \cite{Markiewicz} but additionally
our $\Delta_{CT}$ is measured at high temperature, where the weakening
of the antiferromagnetic correlations and the band broadening would also
reduce $\Delta_{CT}$. Furthermore, polaron effects are likely to be
important in insulating cuprates,\cite{Kastner,Shen} and these effects
would naturally reduce the gap at high temperature.\cite{Kastner} Hence,
one may conclude that the CT gap manifested in the Hall effect at $x$ =
0 is fully consistent with the optics data.

\subsection{Lightly-Doped Region}

\begin{figure}
\includegraphics[clip,width=8.5cm]{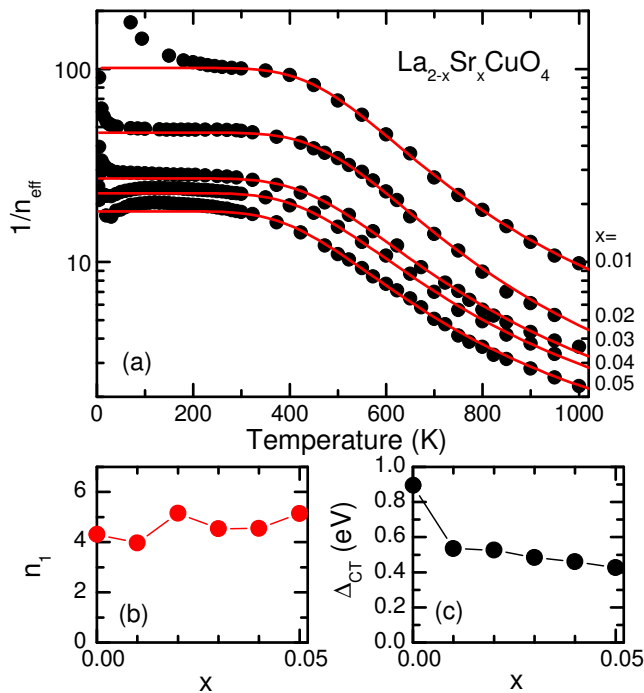} 
\caption{(color online) 
(a) $T$ dependences of $1/n_{\rm eff}$ (= $eR_H/V_{Cu}$) for a series of
LSCO single crystals in the lightly-doped region, $x$ = 0.01 -- 0.05,
with their fits (solid lines) using Eq. (2). The fitting parameters,
$n_1$ and $\Delta_{CT}$, are shown in panels (b) and (c).} 
\end{figure} 

Now that $R_H(T)$ of La$_2$CuO$_4$ is quantitatively understood, let us
see how we can extend this understanding to the lightly-doped region.
Figure 3(a) shows the $R_H(T)$ data of LSCO for $x$ = 0.01 -- 0.05 in a
semi-log plot, where $R_H$ [cm$^3$/C] is converted into the inverse of
effective carrier number per Cu, $1/n_{\rm eff}$ (= $eR_H/V_{Cu}$, which
is nondimensional), for the ease of understanding the meaning of the
numbers. Since the plateau in $R_H(T)$ at moderate temperature gives
$n_{\rm eff}$ that is essentially equal to $x$ at low doping,\cite{Ando}
the impurity term $n_0 e^{-\Delta_{imp}/2k_BT}$ in Eq. (1) should be
replaced with $x$ to describe $R_H(T)$ in this region. \cite{note_terms} 
Hence, we fit the data for $x$ = 0.01 -- 0.05 to 
\begin{equation} 
R_H(T) =
\frac{V_{Cu}}{e} \left( x + n_1 e^{-\Delta_{CT}/2k_{B}T} \right)^{-1} .
\end{equation} 
The solid lines in Fig. 3(a) are the results of the fittings.
\cite{note_x} Note that the upturn at very low temperature seen in all
the data is due to the strong Anderson localization \cite{Ando} that
reduces the number of mobile carriers and naturally causes a deviation
from Eq. (2). Obviously, Eq. (2) gives a reasonable account of the
essential feature of the data (except for the Anderson localization),
and hence one may conclude that the thermal activation of holes gives
rise to the exponential decrease in $R_H$ at high temperature not only
at $x$ = 0 but also at low doping. This in turn indicates that there are
strong charge fluctuations in lightly-doped cuprates at $\gtrsim$400 K,
where the charge transport must become incoherent; therefore, it is
probably not reasonable to describe $R_H$ in this regime using theories
developed for a metallic system (i.e., for coherent electrons with
well-defined wave vectors), such as that in Ref. \onlinecite{Kontani}.

The doping dependences of the parameters $n_1$ and $\Delta_{CT}$ in Eq.
(2) obtained from the fits are shown in Figs. 3(b) and 3(c). It is
notable that $n_1$, a rough measure of the number of available states
for thermal activations (but is amplified by various additional effects
\cite{Adler,note_intrinsic}) is essentially doping-independent for $x$ =
0 -- 0.05 [Fig. 3(b)], which would imply that thermal creations of
carriers of essentially the same nature are taking place in this doping
range. On the other hand, the gap $\Delta_{CT}$ for the thermal
activation [Fig. 3(c)] shows a sudden drop from 0.89 to 0.53 eV upon
doping only 1\% of holes to the parent insulator, but then shows only a
small decrease with $x$. Probably, there are two possibilities to
interpret this result. One is to take the reduction in $\Delta_{CT}$ to
be a result of the softening of the main CT gap upon slight doping; in
this case, the same bands are involved in the activation process after
the doping, and our observation that $n_1$ is essentially doping
independent is in good accord. Considering the fact that doping to a
Mott insulator necessarily involves a change in the electronic structure
at a high energy scale of the order of the on-site repulsion $U$
(because doping one hole to a Mott insulator not only creates a hole
state but also removes one state from the upper Hubbard
band),\cite{Meinders} it would be possible that a slight doping induces
a relatively large change in the band structure. The other possibility
is that the so-called ``in-gap states" \cite{ARPES} are created in the
middle of the original CT gap upon hole doping and our $\Delta_{CT}$
actually measures the charge-transfer excitations from these new states
to the upper Hubbard band (conduction band). In this case, one would
expect $n_1$ for $x \ge$ 0.01 to be much smaller than that for $x$ =0;
however, a large $n_1$ might be possible for some particular shape of
the band edge,\cite{note_n1} so our result in Fig. 3(b) cannot
conclusively exclude this possibility. In any case, the true nature of
$\Delta_{CT}$ in the doped system is best left as an open question, and
its identification is actually at the heart of understanding what really
happens upon doping to a Mott insulator. It is intriguing to note that
our $\Delta_{CT}$ for the lightly-doped region coincides rather well
with the peak frequency of the mid-infrared (MIR) absorption seen in the
optical conductivity of LSCO,\cite{Padilla} so the MIR absorption may
also have something to do with the CT excitations.

In passing, previous studies of the doping dependence of the CT gap
using high-energy probes \cite{Fink,Hill} have found a hardening of the
gap, which appears to be at odds with the first possibility discussed
above. However, Markiewicz and Bansil argued \cite{Markiewicz} that
those high-energy experiments may only see hard branches of the various
modes of the CT excitations; naturally, our thermodynamic measurement
probes the CT excitation of the lowest energy, which may not be easily
seen by the high-energy probes. In this regard, it should be noted that
our $\Delta_{CT}$ measures the effective excitation energy {\it at high
temperature}, which is naturally smaller than the band gap at $T$ = 0,
so a care must be taken when comparing our $\Delta_{CT}$ to that
calculated theoretically for $T$ = 0.

\subsection{Superconducting Doping Range}

\begin{figure}
\includegraphics[clip,width=8cm]{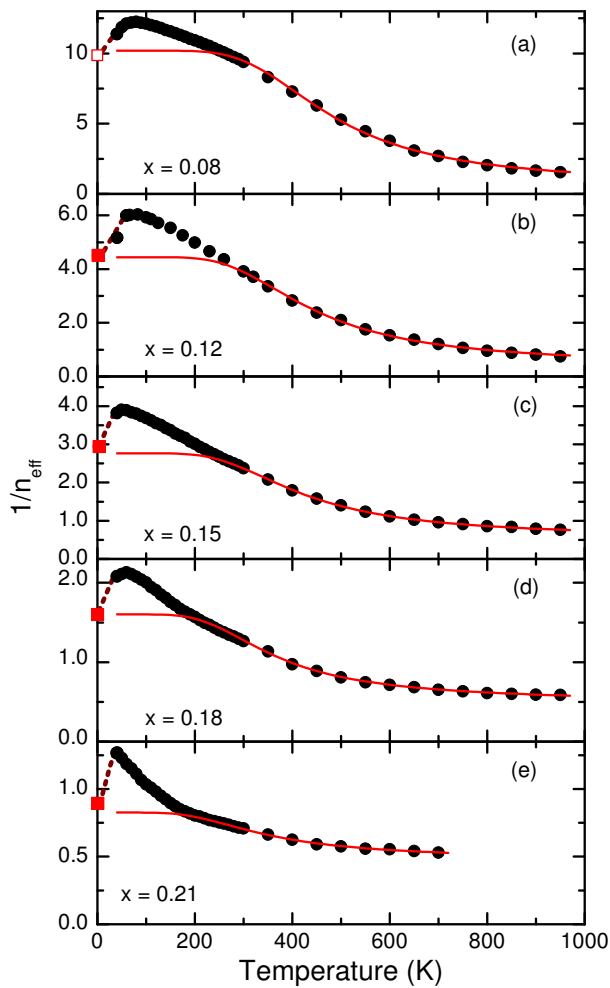} 
\caption{(color online) 
The $R_H(T)$ data, converted into the nondimensional $1/n_{\rm eff}$,
for $x$ = 0.08 -- 0.21 and their fits (solid lines) using Eq. (4). The
squares at $T$ = 0 mark $R_H(0)$ obtained from the pulsed magnetic-field
experiments:\cite{Balakirev2} The $R_H(0)$ value for $x$ = 0.08 is
obtained by extrapolating the data in Ref. \onlinecite{Balakirev2} for
10 -- 40 K to $T \rightarrow$ 0; for $x$ = 0.15, 0.18, and 0.21, the
$R_H(0)$ values are obtained by a simple interpolation of those for
$x$ = 0.14, 0.16, 0.17, 0.20, and 0.22 reported in Ref.
\onlinecite{Balakirev2}. The dashed lines are guides to the eyes.}
\end{figure} 

At higher doping $x \ge$ 0.08, Eq. (2) becomes obviously inappropriate,
because $n_{\rm eff}$ at moderate temperature is no longer equal to $x$.
\cite{Ando} Also, the plateau in $R_H(T)$ at moderate temperature
observed in the lightly-doped region is now replaced by a peaked
temperature dependence, which points to the necessity of an elaborate
model to understand its behavior at low to moderate temperature.
Nevertheless, as is obvious in Fig. 1, in those samples the behavior of
$R_H(T)$ at high temperature is surprisingly similar to that observed in
lightly-doped samples and is changing only gradually with doping; this
observation strongly suggests that essentially the same thermal
activation process is affecting the $R_H(T)$ behavior at high
temperature even in the superconducting doping range. \cite{note_GT}
Motivated by this qualitative observation, we attempt to understand the
high-temperature $R_H(T)$ behavior for $x \ge$ 0.08 in terms of a crude
phenomenological two-carrier model, remembering that the occurrence of
an electronic heterogeneity (microscopic phase separation or coexistence
of different types of carriers) has been discussed repeatedly for LSCO;
\cite{Orenstein,Muller,Uemura,Gorkov,Mayr,Kivelson,Tranquada} for
example, doping evolutions of the magnetic susceptibility \cite{Muller}
or the superfluid density \cite{Uemura} have been discussed to reflect
an electronic heterogeneity. In the following analysis, we crudely
hypothesize that there are two types of holes, those that live on the
Fermi arc (or small hole pockets), and others that live on a large Fermi
surface (FS). We suppose that the former contribute to the Hall effect
with the component $R_H^{arc}$ described by Eq. (2), while the latter
contribute with a component $R_H^{LFS} = V_{Cu}/en_2$. For the sake of
simplicity, we further suppose that these two components $R_H^{arc}$ and
$R_H^{LFS}$ are additive in the expression of the total Hall coefficient
with their respective fractions $f$ and $1-f$, as is the case with the
two-carrier model proposed for cuprates by Lee and
Nagaosa.\cite{Lee-Nagaosa} Hence, we try to see how $R_H(T)$ at high
temperature can be described in terms of the expression
\begin{eqnarray} 
R_H(T) = f R_H^{arc} + (1-f) R_H^{LFS} \\
 = \frac{V_{Cu}}{e} \left[ f/(x + n_1 e^{-\Delta_{CT}/2k_{B}T})
+(1-f)/n_2 \right] .
\end{eqnarray} 
We emphasize that Eq. (4) should be considered to be essentially a
working hypothesis that is not backed by a concrete theory, and this
expression naturally fails at low to moderate temperature where $R_H$
shows a nontrivial temperature dependence due, for example, to the
scattering-time anisotropy.\cite{Stojkovic,Ong} Actually, already in the
upper end of the lightly-doped region discussed in the previous
subsection (i.e., $x$ = 0.04 and 0.05), a slight deviation of the data
from the fitting is evident below 300 K in Fig. 3(a), which is likely to
be due to the intrinsic temperature dependence of $R_H$ in the coherent
regime.

The solid lines in Figs. 4(a)-4(e) show the fits of the data for $x$ =
0.08 -- 0.21 to Eq. (4). It is rather surprising that our crude model
very well describes the experimental data for $T > 300$ K. As is noted
above, it is expected from the beginning that Eq. (4) does not fit the
data at low to moderate temperature, but intriguingly the value of $R_H$
in the plateau region of our fit turned out to bear reasonable physical
meaning: Remember, it has been known from the pulsed high-magnetic-field
experiments \cite{Balakirev,AndoHall,Balakirev2} that the normal-state
Hall coefficient in the $T \rightarrow 0$ limit, $R_H(0)$, which is
expected to genuinely reflect the electronic structure,\cite{Ong} is
much smaller than that at $T_c$. We plot $R_H(0)$ obtained from recent
measurements \cite{Balakirev2} of LSCO with squares in Figs. 4(a)-4(e),
which agree reasonably well with what our fittings would
suggest.\cite{note_fitting} This gives phenomenological support to the
validity of Eq. (4), and implies that the evolution of the effective
carrier density at low temperature, which would be directly reflected in
the measured $R_H$ for $T \rightarrow$ 0, may be understood in the
framework of a two-carrier model in LSCO. The apparent success of our
crude model in describing the $R_H$ values at both high temperature and
at 0 K at the same time seems to suggest that there is a certain truth
in the above approach and calls for theoretical attentions.

\begin{figure}
\includegraphics[clip,width=8.5cm]{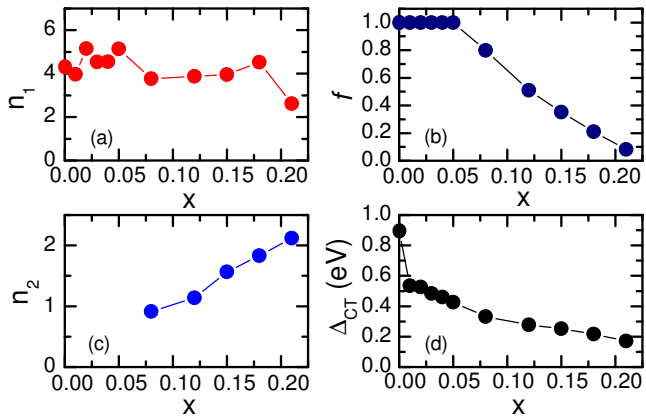} 
\caption{(color online) 
Doping dependences of the fitting parameters for $R_H(T)$ of LSCO. 
(a) $n_1$, a rough measure of the number of available states in the 
bands relevant to the thermal activation in the component $R_H^{arc}$ 
which is presumably associated with the Fermi arc; 
(b) $f$, fraction of the component $R_H^{arc}$; 
(c) $n_2$, effective hole number in the component $R_H^{LFS}$ 
presumably associated with a large Fermi surface; 
(d) $\Delta_{CT}$, energy gap for the thermal activation.} 
\end{figure} 

The doping dependences of the fitting parameters are shown in Fig. 5, in
which the results for the parent insulator and the lightly-doped samples
discussed in the previous subsections are combined. It is notable that
$n_1$ is essentially doping-independent up to $x$ = 0.18 [Fig. 5(a)],
which suggests that the physical origin of the thermally activated
contribution is essentially unchanged up to optimum doping. On the other
hand, the fraction $f$ for the component $R_H^{arc}$, only in which the
thermal activation is relevant, decreases rapidly for $x >$ 0.05 [Fig.
5(b)], which causes the overall temperature dependence in $R_H$ to
become weaker as the overdoped region is approached. The parameter
$n_2$, the effective hole number associated with the large FS, is of the
order of 1 as expected [Fig. 5(c)], and its doping dependence is likely
to reflect the change in the FS shape observed by angle-resolved
photo-emission spectroscopy (ARPES) experiments;\cite{ARPES} namely, as
the shape of the large FS changes from hole-like to electron-like upon
overdoping, $R_H^{LFS}$ is expected to present a sign change, and hence
$n_2$ (= $V_{Cu}/eR_H^{LFS}$) should eventually diverge.

\begin{figure}
\includegraphics[clip,width=8.5cm]{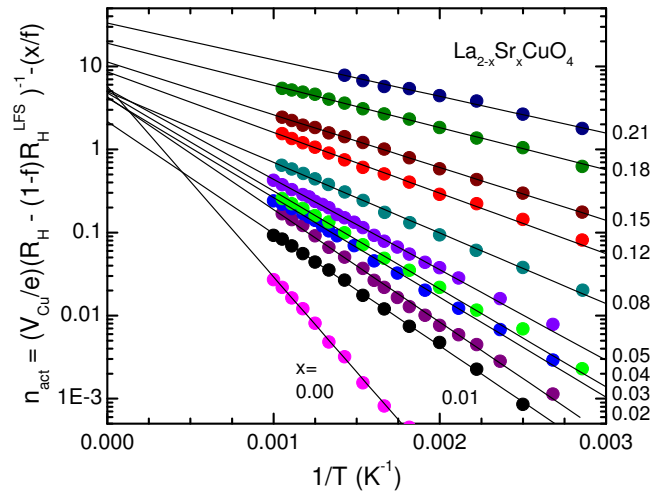} 
\caption{(color online) 
Arrhenius plot of the activation term $n_{\rm act}$, obtained from 
the $R_H$ data by subtracting the constant contributions, for all 
$x$ values. The solid lines are straight-line fits to the data to
emphasize the activated behavior.} 
\end{figure} 

To demonstrate the quality of our analysis, we show in Fig. 6 the
Arrhenius plot of the activation term $n_{\rm act} [= (n_1/f)
e^{-\Delta_{CT}/2k_BT}]$, which is obtained from the $R_H$ data after
subtracting the constant contribution from the $R_H^{LFS}$ component and
that from the $x$-number of holes in the $R_H^{arc}$ component; namely,
after determining all the parameters of Eq. (4),\cite{note_fitting} we
obtain $n_{\rm act}$ by calculating
$[eR_H/V_{Cu}-(1-f)/n_2]^{-1}-(x/f)$. Remember, the constant
contributions used for calculating $n_{\rm act}$ from the raw $R_H$ data
are justified by the reasonable agreement between $R_H(0)$ and our fits
for $T \rightarrow 0$. Figure 6 clearly testifies that the
$T$-dependence of $R_H$ above $\sim$300 K is of activated nature. 

As discussed in Sec. III-B, the gap $\Delta_{CT}$ for the thermal
activation [Fig. 5(d)] shows a sudden drop from $x$ = 0 to 0.01 and then
shows only a small decrease with $x$ in the lightly-doped region where
our analysis for obtaining $\Delta_{CT}$ is robust. This weak doping
dependence of $\Delta_{CT}$ appears to continue into the superconducting
doping range, although one should take this result with a grain of salt,
given the crude nature of our analysis for this region. Nevertheless, it
is probably reasonable to conclude from our analysis that there remain
charge fluctuations associated with some charge-transfer excitations
deep into the superconducting doping region, and such fluctuations keep
causing a thermally-activated temperature dependence in $R_H$ at high
temperature.

\section{DISCUSSIONS}

\begin{figure}
\includegraphics[clip,width=8.5cm]{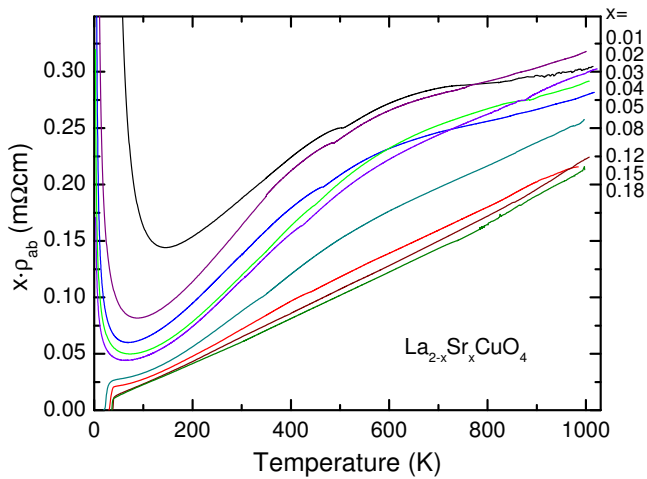} 
\caption{(color online) 
$T$-dependences of the product $x\rho_{ab}$ for $x$ = 0.01 -- 0.18 
up to 1000 K; note that $x$ is a non-dimensional parameter.} 
\end{figure} 

The doping dependence of $f$ shown in Fig. 5(b) suggests that the
crossover in dominance from the Fermi arc to the large FS is pivoted at
$x$ = 0.08, and intriguingly this crossover around $x$ = 0.08 appears to
be reflected in a qualitative change in the temperature dependence of
the in-plane resistivity $\rho_{ab}$. This situation is most easily seen
in the plot of $x\rho_{ab}$ vs. $T$ up to 1000 K shown in Fig. 7; here,
the data for $x$ = 0.01 -- 0.05 show a clear tendency towards saturation
at high temperature, which presumably comes from the increasing number
of thermally-created holes, whereas such a tendency is mostly absent in
the data for $x$ = 0.12 -- 0.18 where the thermal activation occurs only
in the minority component. It is useful to note that, within our model,
the contribution from $R_H^{arc}$ to the total $R_H$ remain relatively
large at low temperature even when $f$ is small because of the smallness
of $x$ compared to $n_2$ [note that $R_H$ at low temperature is written
as $(V_{Cu}/e)\{ f/x + (1-f)/n_2 \}$]; for example, even though $f$
is only 0.2 at $x$ = 0.18, roughly 2/3 of the total $R_H$ is due to the
component $R_H^{arc}$ at low temperature, which is responsible for the
60\% decrease in the total $R_H$ from 200 to 950 K. Hence, it is
possible that $\rho_{ab}$ is insensitive to what is happening in the
minority component, while $R_H$ is more sensitive to the thermal
activations taking place in the minority component.

\begin{figure}
\includegraphics[clip,width=8.5cm]{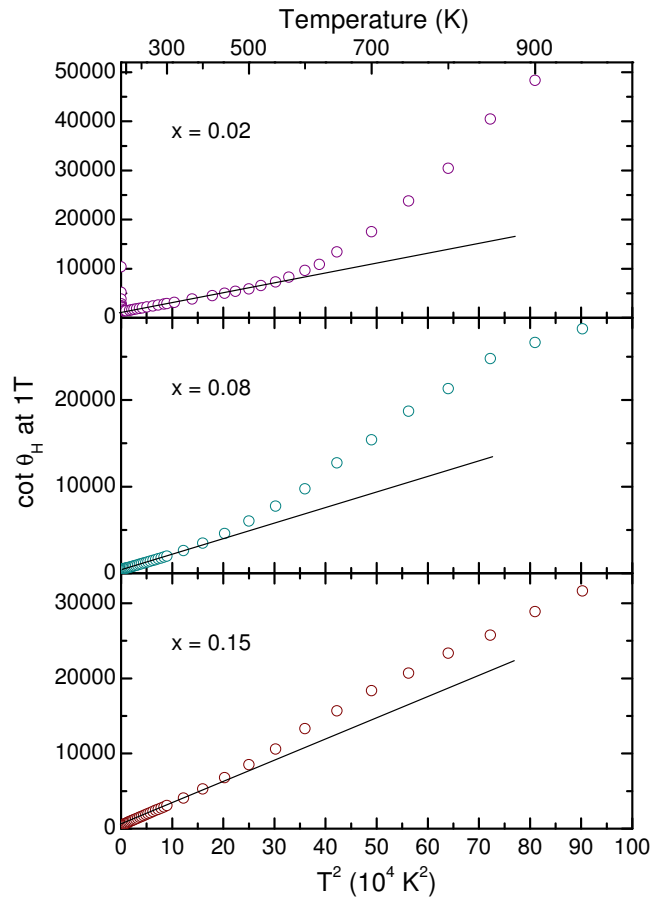} 
\caption{(color online) 
Plots of $\cot \theta_H$ (at 1 T) vs $T^2$ for three representative
dopings. The solid lines emphasize the $T^2$ law that holds at low 
temperature.} 
\end{figure} 

It is useful to mention that in cuprates there is a celebrated
``scattering-time separation" of charge carriers (i.e., the $T$
dependences of resistivity and cotangent of the Hall angle, $\cot
\theta_H$, present different powers of $T$).\cite{Orenstein} In our
data, $\cot \theta_H$ behaves as $\sim T^2$ only below $\sim$400 K in
the superconducting doping range (Fig. 8), which suggests that the
deviation of $R_H$ from the fittings shown in Figs. 4(a)-4(e) may be
linked to the appearance of two scattering times. This is reasonable,
because thermally-created carriers are incoherent and blur the
scattering-time separation at high temperature; also, the
scattering-time separation would disappear at low enough temperature
where the elastic impurity scattering dominates, if it is due to a
development of scattering-time anisotropy in the momentum
space.\cite{Stojkovic,Ong,Ioffe,Hussey} Hence, our data seem to suggest
that the scattering-time separation is a result of a peculiar
scattering-time anisotropy,\cite{Stojkovic,Ioffe,Hussey} which develops
only at intermediate temperature. In passing, similar scattering-time
separations have been found in non-cuprate materials,
\cite{Paschen,Yeh,Nakajima,Rosenbaum} suggesting that this phenomenon
may not be very special.

It is worth emphasizing that the picture derived here (appearance of
activated carriers at high temperature and growth of the contribution
from a large FS in the superconducting doping range) not only explains
the Hall effect but also is consistent with the resistivity data, as can
be inferred in Fig. 7. Also, the magnetic susceptibility data up to 800
K (Ref. \onlinecite{Johnston}) have been analyzed \cite{Muller} with a
similar model and are essentially consistent with our conclusion. In
view of the present picture, we can probably address some of the
complications regarding the pseudogap.\cite{Orenstein} In the past, a
scaling of $R_H(T)$ was used \cite{Hwang} to identity a characteristic
temperature $T^*$ that reaches $\sim$700 K at low doping, and this $T^*$
was shown to agree with those deduced from the dc magnetic
susceptibility or the Knight shift. Since this $T^*$ is much higher than
the pseudogap temperature deduced from spectroscopic
means,\cite{Orenstein} it is often called ``large pseudogap". Our
results imply that the large pseudogap is related to the freezing of the
thermal creation of charge carriers, which causes the chemical potential
to move and thereby affects various physical quantities.

\section{CONCLUSION}

We have shown that the behavior of $R_H(T)$ in insulating LSCO ($x$ = 0
-- 0.05) at high temperature clearly signifies that charge carriers are
thermally activated over a sub-eV gap, likely associated with some
charge-transfer excitations. In the superconducting doping region, the
high-temperature $R_H(T)$ behavior is found to be qualitatively the same
but is weakened, which we try to describe with a phenomenological
two-carrier model that considers the thermal activation for one of the
two components. Besides giving a perspective to consistently understand
the high-temperature behavior of $R_H$ and $\rho_{ab}$, our model allows
us to understand the $R_H$ value both at high temperature and at 0 K for
a wide range of doping, despite its crudeness. Overall, our data and
analysis strongly suggest that charge fluctuations associated with a
sub-eV gap should be taken into account when discussing the physics of
cuprates even at relatively low energy scales. Implications of this
conclusion include: (1) breakdown of the $T^2$ law of $\cot \theta_H$ at
high temperature may be related to the onset of incoherence in one of
the two species of the carriers, that dominates the Hall effect but
little influences the resistivity near optimum doping; (2) ``opening of
the large pseudogap" may actually be a ``freezing of charge
fluctuations" associated with the Fermi arc.

\begin{acknowledgments} 

We greatly thank L. P. Gor'kov and G. B. Teitel'baum for illuminating
discussions and pointing out an activated behavior in our data. We also
thank E. Abrahams, B. Batlogg, T. H. Geballe, J. Fink, S. A. Kivelson,
P. B. Littlewood, N. P. Ong, D. Pines, G. A. Sawatzky, X. F. Sun, A. A.
Taskin, T. Tohyama, I. Tsukada, and S. Uchida for helpful discussions.
This work was supported by Grant-in-Aid for Science from JSPS.

\end{acknowledgments}


\begin{thebibliography}{} 

\bibitem{Balakirev}
F. F. Balakirev {\it et al.}, 
Nature {\bf 424}, 912 (2003).

\bibitem{Paschen}
S. Paschen {\it et al.}, 
Nature {\bf 432}, 881 (2004).

\bibitem{Dagan}
Y. Dagan, M. M. Qazilbash, C. P. Hill, V. N. Kulkarni, and R. L. Greene,
Phys. Rev. Lett. {\bf 92}, 167001 (2004).

\bibitem{Yeh}
A. Yeh  {\it et al.}, 
Nature {\bf 419}, 459 (2002).

\bibitem{Noda}
T. Noda, H. Eisaki, and S. Uchida,
Science {\bf 286}, 265 (1999).

\bibitem{Orenstein}
J. Orenstein and A. Millis, 
Science {\bf 288}, 468 (2000).

\bibitem{Kivelson1}
S. A. Kivelson {\it et al.}, 
Rev. Mod. Phys. {\bf 75}, 1201 (2003).

\bibitem{Anderson}
P. W. Anderson, Phys. Rev. Lett. {\bf 67}, 2092 (1991).

\bibitem{Stojkovic}
B. P. Stojkovic and D. Pines, 
Phys. Rev. B {\bf 55}, 8576 (1997).

\bibitem{Kontani}
H. Kontani, K. Kanki, and K. Ueda, 
Phys. Rev. B {\bf 59}, 14723 (1999).

\bibitem{Veberic}
D. Veberic and P. Prelovsek, 
Phys. Rev. B {\bf 66}, 020408(R) (1999).

\bibitem{ARPES}
A. Damascelli, Z. Hussain, and Z.-X. Shen, 
Rev. Mod. Phys. {\bf 75}, 473 (2003).

\bibitem{Ando}
Y. Ando, Y. Kurita, S. Komiya, S. Ono, and K. Segawa, 
Phys. Rev. Lett. {\bf 92}, 197001 (2004).

\bibitem{Kastner}
M. A. Kastner, R. J. Birgeneau, G. Shirane, and Y. Endoh, 
Rev. Mod. Phys. {\bf 70}, 897 (1998).

\bibitem{Dagotto}
E. Dagotto, 
Rev. Mod. Phys. {\bf 66}, 763 (1994).

\bibitem{Geballe}
T. H. Geballe and B. Y. Moyzhes, 
Ann. Phys. (Leipzig) {\bf 13}, 20 (2004).

\bibitem{Komiya1}
S. Komiya, Y. Ando, X. F. Sun, and A. N. Lavrov, 
Phys. Rev. B {\bf 65}, 214535 (2002).

\bibitem{Komiya2}
S. Komiya, H.-D. Chen, S.-C. Zhang, and Y. Ando, 
Phys. Rev. Lett. {\bf 94}, 207004 (2005).

\bibitem{Taskin_APL}
A. A. Taskin, A. N. Lavrov, and Y. Ando, 
Appl. Phys. Lett. {\bf 86}, 091910 (2005).

\bibitem{Kanai}
H. Kanai {\it et al.}, 
J. Solid State Chem. {\bf 131}, 150 (1997).

\bibitem{Nishikawa}
T. Nishikawa, J. Takeda, and M. Sato, 
J. Phys. Soc. Jpn. {\bf 63}, 1441 (1994).

\bibitem{Hwang}
H. Y. Hwang {\it et al.}, 
Phys. Rev. Lett. {\bf 72}, 2636 (1994).

\bibitem{note_imp} 
The binding energy of the holes to the oxygen acceptor mentioned in Ref.
\onlinecite{Kastner} is the activation energy obtained from the Arrhenius
plot, which corresponds to 1/2 of $\Delta_{imp}$ defined in Eq. (1).
(This is because the chemical potential moves to the middle of the gap
when the thermal activation process becomes dominant). Hence, our result
of $\Delta_{imp}$ = 87 meV is essentially consistent with Ref.
\onlinecite{Kastner} where the binding energy was mentioned to be 35
meV.

\bibitem{Adler}
D. Adler and J. Feinleib,
Phys. Rev. B {\bf 2}, 3112 (1970).

\bibitem{note_intrinsic}
Remember that in an intrinsic semiconductor the Hall coefficient is
given by $R_H = (\mu_h - \mu_e)/[ne(\mu_h + \mu_e)]$, where $\mu_h$
($\mu_e$) is the mobility of holes (electrons) and $n = n_1
e^{-\Delta/2k_BT}$ is the number of pair-created electrons and holes.
Hence, the effective density of states is always enhanced by the factor
$(\mu_h+\mu_e)/|\mu_h - \mu_e|$. 

\bibitem{Markiewicz}
R. S. Markiewicz and A. Bansil,
Phys. Rev. Lett. {\bf 96}, 107005 (2006).

\bibitem{Shen}
K. M. Shen {\it et al.}, 
Phys. Rev. Lett. {\bf 93}, 267002 (2004).

\bibitem{note_terms}
In lightly Sr-doped samples, the impurity states of the order of 0.5\%
estimated for $x$ = 0 are probably still present, but they provide only
a small amount of mobile holes at moderate temperature (e.g., 0.1\% at
300 K) because $\Delta_{imp}$ = 87 meV = 1010 K; hence, the impurity
term can be safely neglected in Eq. (2) in comparison with $x$.

\bibitem{note_x}
In those fittings, $x$ was actually treated as a fitting parameter,
because there is an uncertainty of up to 10\% in the compositional $x$
values in lightly-doped samples. The resulting $x$ values obtained from
the fits are 0.0098, 0.0213, 0.0369, 0.0440, and 0.0546 for the five
sets of data, all in reasonable agreement with the compositional values.

\bibitem{Meinders}
M. B. J. Meinders, H. Eskes, and G. A. Sawatzky, 
Phys. Rev. B {\bf 48}, 3916 (1993).

\bibitem{note_n1}
In conventional semiconductor physics, $n_1$ is actually a
temperature-dependent effective density of states that is viewed to be
concentrated at the band edge, and its value is determined largely by
the shape of the actual density of states near the edge. The
temperature dependence of $n_1$ (which, for example, is $T^{2/3}$ in the
case of a parabolic band in a 3D system) is often neglected because it
is much weaker than the exponential term.

\bibitem{Padilla}
W. J. Padilla, Y. S. Lee, M. Dumm, G. Blumberg, S. Ono, K. Segawa,
S. Komiya, Y. Ando, and D. N. Basov,
Phys. Rev. B {\bf 72}, 060511(R) (2005).

\bibitem{Fink}
J. Fink {\it et al.}, 
J. Electron Spectroscopy and Related Phenomena {\bf 66}, 395 (1994).

\bibitem{Hill}
Y.-J. Kim {\it et al.}, 
Phys. Rev. B {\bf 70}, 094524 (2004).

\bibitem{note_GT}
We note that recently Gor'kov and Teitel'baum have tried to fit the
high-temperature $R_H(T)$ data for the whole doping range by simply
using Eq. (2) with the first term replaced by $n_0(x)$; L. P. Gor'kov
and G. B. Teitel'baum, cond-mat/0607010.

\bibitem{Muller}
K. A. M\"{u}ller, 
Physica C {\bf 341-348}, 11 (2000).

\bibitem{Uemura}
Y. J. Uemura, 
Solid State Commun. {\bf 120}, 347 (2001).

\bibitem{Gorkov}
L. P. Gor'kov, 
J. Supercond. {\bf 14}, 365 (2001).

\bibitem{Mayr}
M. Mayr, G. Alvarez, A. Moreo, and E. Dagotto,
Phys. Rev. B {\bf 73}, 014509 (2006).

\bibitem{Kivelson}
S. A. Kivelson, E. Fradkin, and T. H. Geballe,
Phys. Rev. B {\bf 69}, 144505 (2004).

\bibitem{Tranquada}
J. M. Tranquada, Proc. SPIE {\bf 5932}, 59320C (2005).

\bibitem{Lee-Nagaosa}
P. A. Lee and N. Nagaosa, 
Phys. Rev. B {\bf 46}, 5621 (1992).

\bibitem{Ong}
N. P. Ong,
Phys. Rev. B {\bf 43}, 193-201 (1991).

\bibitem{AndoHall}
Y. Ando {\it et al.}, 
Phys. Rev. B {\bf 56}, R8530 (1997).

\bibitem{Balakirev2}
F. F. Balakirev, J. B. Betts, A. Migliori, G. S. Boebinger,
I. Tsukada, and Y. Ando, NHMFL 2005 Annual Report.

\bibitem{note_fitting}
In our fitting procedure, the parameters are determined so that the
widest range of the high-temperature data are fitted to the exponential 
behavior. Hence, the temperature below which the fitting fails and the 
the plateau value of our fit come out naturally without any subjective 
choice in the procedure.

\bibitem{Ioffe}
L. B. Ioffe and A. J. Millis, Phys. Rev. B {\bf 58}, 11631 (1998).

\bibitem{Hussey}
N. E. Hussey, Eur. Phys. J. B {\bf 31}, 495 (2003).

\bibitem{Nakajima}
Y. Nakajima {\it et al.}, J. Phys. Soc. Jpn. {\bf 73}, 5 (2004).

\bibitem{Rosenbaum}
T. F. Rosenbaum, A. Husmann, S. A. Carter, and J. M. Honig,
Phys. Rev. B {\bf 57}, R13997 (1998).

\bibitem{Johnston}
D. C. Johnston, Phys. Rev. Lett. {\bf 62}, 957 (1989).

\end{thebibliography}
\end{document}